\documentclass[11pt]{article}

\usepackage[utf8]{inputenc}
\usepackage[T1]{fontenc}
\usepackage{lmodern}
\usepackage{microtype}
\usepackage{booktabs}
\usepackage{amsmath}
\usepackage{amssymb}
\usepackage{graphicx}
\usepackage[margin=1in]{geometry}
\usepackage{enumitem}
\usepackage{xcolor}
\usepackage{url}
\usepackage{caption}
\usepackage{hyperref}

\hypersetup{
  colorlinks=true,
  linkcolor=blue!45!black,
  citecolor=blue!45!black,
  urlcolor=blue!45!black,
  pdftitle={Queen-Bee Agents: A BeeSpec-Centered Architecture for Governed Enterprise MCP Orchestration},
  pdfauthor={Dutao Zhang and Liaotian}
}

\setlist{nosep,leftmargin=1.5em}
\newcommand{\keywords}[1]{\vspace{0.75em}\noindent\textbf{Keywords:} #1}

\title{Queen-Bee Agents: A BeeSpec-Centered Architecture for Governed Enterprise MCP Orchestration}
\author{
Dutao Zhang\thanks{Corresponding author.} and Liaotian\\
\small Macao Polytechnic University\\
\small \texttt{p2527418@mpu.edu.mo}
}
\date{\small Technical Report / arXiv Preprint}

\begin{document}

\maketitle

\begin{abstract}
Enterprise agent systems increasingly need to connect large language models to private tools, internal knowledge, and Model Context Protocol (MCP) interfaces. In this setting, raw task capability is insufficient: organizations also require policy enforcement, tenant-scoped isolation, and execution that remains within explicit operational boundaries. We present \textit{Queen-Bee}, a governed multi-agent architecture in which a Queen control plane retrieves capabilities, plans task-scoped execution, and compiles a structured \textit{BeeSpec} that is executed by specialized Bee agents under constrained tool access. We implement a working prototype with tenant-scoped MCP connectors, audit-backed execution-time governance, retrieval-driven weak incubation, and multiple provisioning backends. We evaluate the system on 59 enterprise-style tasks spanning governance-sensitive requests, retrieval-driven provisioning, scoped local execution, and chemistry workflow integration. The retrieval-driven Queen-Bee variant achieves a task success rate of 0.964, zero governance failures, and substantially better scoped execution quality than both a static Queen-Bee baseline and a permissive single-agent baseline. We further show a multi-Bee chemistry workflow with explicit approval gating and a concrete top-3 shortlist grounded in real upstream evidence and screening artifacts. Additional comparisons with hybrid retrieval and LLM-guided provisioning show that richer provisioning backends are viable but do not outperform the lightweight structured retriever on the current small, highly structured capability registry. The results provide prototype-level systems evidence rather than a production deployment study, and suggest that enterprise agent platforms should be evaluated not only by capability, but also by governed provisioning, isolation behavior, scoped execution quality, and artifact-aware workflow coordination.
\end{abstract}


\section{Introduction}

Recent LLM-based agent systems can invoke tools, access structured resources, and execute workflows across external systems. However, enterprise deployment introduces a substantially different problem from open-domain agent evaluation. In enterprise private MCP environments, an agent must not only complete tasks but do so while respecting department boundaries, tenant isolation, tool governance, and audit requirements. A system that completes a task by invoking the wrong tool or crossing tenant boundaries is operationally unacceptable even if the end result appears useful.

This observation motivates a shift away from monolithic, broadly privileged agents toward governed, role-constrained orchestration. We argue that enterprises do not primarily need a single general agent with unrestricted tool access. Instead, they need a system that can provision specialized execution units, assign bounded capabilities, and enforce execution-time policy constraints.

To support this view, we introduce \textit{Queen-Bee}, a governed multi-agent architecture for enterprise MCP integration. The central design decision is to separate planning and execution through a structured intermediate representation, \textit{BeeSpec}. Rather than sending a request directly to an execution agent, the Queen first retrieves relevant capabilities, plans the execution boundary, and compiles a BeeSpec containing the role, attached skills, allowed MCP-backed tools, tenant scope, memory scope, and policy profile for a Bee. Execution then proceeds under those constraints.

The paper makes four contributions:
\begin{enumerate}[leftmargin=1.5em]
\item We present a BeeSpec-centered architecture that separates the enterprise agent control plane from the execution plane.
\item We implement a governed prototype with tenant-scoped MCP connectors, real stdio MCP adapters, retrieval-driven weak incubation, and audit-backed execution-time policy checks.
\item We design a three-part evaluation covering isolation and governance, retrieval-driven provisioning, and scoped Bee execution over 59 enterprise-style tasks.
\item We compare multiple provisioning backends, including lightweight structured retrieval, noisy-registry stress retrieval, hybrid sparse+dense retrieval, and LLM-guided provisioning, and show that the lightweight structured backend remains strongest under the current registry scale.
\end{enumerate}

\section{Background and Motivation}

Enterprise requests are rarely single-step or single-boundary tasks. Recruiting workflows may require candidate lookup, scheduling, and policy-sensitive restrictions around compensation data. IT workflows may require knowledge lookup, ticket creation, and clear separation from HR systems. Cross-tenant requests must be rejected even when the apparent operation is otherwise benign. These constraints expose a structural weakness in single-agent designs: broad capability is often coupled with broad privilege.

Existing work on tool use and multi-agent coordination demonstrates that language models can act over external environments and decompose tasks across roles \cite{schick2023toolformer,yao2023react,wu2023autogen,li2023camel}. Yet enterprise deployment places additional emphasis on bounded execution, auditability, and tenant-aware connectivity. Our position is therefore not that multi-agent systems are universally better, but that enterprise MCP environments require an architecture in which orchestration, capability assignment, and execution-time governance are explicit first-class components.

\section{Related Work}

Recent LLM agent work has shown strong progress in tool use, reasoning-action coupling, and multi-agent role decomposition. Toolformer and ReAct illustrate two complementary directions: learning or prompting models to use tools while interleaving reasoning and acting \cite{schick2023toolformer,yao2023react}. ToolLLM extends this trend toward large API ecologies, emphasizing broad tool coverage rather than bounded enterprise execution \cite{qin2023toolllm}. Multi-agent frameworks such as AutoGen and CAMEL show that role-specialized agents can coordinate over complex tasks \cite{wu2023autogen,li2023camel}, but they do not directly solve tenant isolation, tool governance, or explicit execution-boundary compilation.

Our work is closer in spirit to software architecture and access-control traditions than to open-ended agent societies. The governance layer draws on role-based access control and enforceable policy thinking \cite{sandhu1996rbac,schneider2000enforce}, while the BeeSpec abstraction explicitly separates control-plane planning from execution-plane action. The goal is not to maximize the number of tools an agent can reach, but to make capability assignment, auditing, and stage-level workflow coordination explicit and inspectable.

\section{Queen-Bee Architecture}

\subsection{Overview}

Queen-Bee separates execution into four layers:
\begin{enumerate}[leftmargin=1.5em]
\item \textbf{Queen control plane}, responsible for capability retrieval, blueprint planning, BeeSpec generation, and governance decisions.
\item \textbf{BeeSpec intermediate layer}, which explicitly defines the execution boundary for each Bee.
\item \textbf{Bee execution plane}, in which specialized Bees execute only within the capabilities assigned by BeeSpec.
\item \textbf{Tenant-scoped MCP connector layer}, which resolves tool invocations inside the active tenant scope.
\end{enumerate}

\subsection{BeeSpec as the Core Intermediate Layer}

BeeSpec is the central architectural interface between planning and execution. It contains the Bee identity, role, domain, attached skills, allowed tools, policy profile, memory scope, and tenant scope. This layer matters because it decouples provisioning from execution. The Queen decides what a Bee is allowed to do; the Bee runtime carries out the task under those constraints. This separation makes routing, auditing, and policy enforcement much easier to reason about than in a monolithic agent.

Concretely, the prototype represents each BeeSpec as a structured record with the following fields:

\begin{table}[t]
\centering
\small
\setlength{\tabcolsep}{5pt}
\begin{tabular}{ll}
\toprule
Field & Meaning \\
\midrule
\texttt{bee\_id} & Unique execution-unit identifier for audit and trace linking \\
\texttt{role} & Natural-language role assigned by the Queen \\
\texttt{domain} & Operational domain, such as HR, IT, finance-sensitive, or chemistry \\
\texttt{tenant\_scope} & Tenant boundary within which MCP calls must execute \\
\texttt{memory\_scope} & Accessible memory namespace for the Bee \\
\texttt{attached\_skills} & Retrieved skills that drive local execution planning \\
\texttt{allowed\_tools} & MCP-backed tools authorized for this Bee \\
\texttt{policy\_profile} & Guardrail profile used before each tool invocation \\
\texttt{approval\_gate} & Optional human approval requirement before downstream execution \\
\bottomrule
\end{tabular}
\caption{BeeSpec schema used as the intermediate representation between Queen planning and Bee execution.}
\label{tab:beespec-schema}
\end{table}

\subsection{Queen Responsibilities}

The Queen is not a general execution agent. It operates at the control plane and performs four functions:
\begin{enumerate}[leftmargin=1.5em]
\item capability retrieval over MCP and skill registries,
\item blueprint planning for task-scoped execution,
\item BeeSpec generation and tenant-scoped Bee provisioning,
\item execution-time governance through policy checks and audit logging.
\end{enumerate}

\subsection{Bee Responsibilities}

Each Bee is a specialized execution unit that operates under BeeSpec constraints. In the present prototype, Bees are domain-scoped to HR and IT. Bees use attached skills to derive local tool plans and invoke tenant-scoped MCP-backed tools only after authorization by the policy layer.

\section{Prototype}

We implement a Python prototype with domain-scoped enterprise Bees, tenant-scoped MCP connectors, and a policy engine that mediates every tool invocation. The connector layer uses real stdio MCP adapters backed by a local FastMCP server, together with persistent tenant-scoped session reuse. This keeps the prototype closer to real tool invocation than a mock in-process tool registry while still allowing controlled evaluation.

The implementation is intended as a prototype and evaluation harness, not as a claim of production-grade enterprise security. Its purpose is to make the control-plane, BeeSpec, connector, and policy boundaries executable enough to measure routing, provisioning, isolation, and workflow behavior.

The system includes two capability registries:
\begin{itemize}[leftmargin=1.5em]
\item an \textbf{MCP registry}, containing tool descriptions, domains, risk levels, and tenant scope metadata;
\item a \textbf{skill registry}, containing reusable execution skills with required tool dependencies.
\end{itemize}

The Queen can provision Bees through four backend styles:
\begin{enumerate}[leftmargin=1.5em]
\item \textbf{lightweight structured retrieval},
\item \textbf{stress retrieval} under a noisier registry,
\item \textbf{hybrid sparse+dense retrieval},
\item \textbf{LLM-guided provisioning} over retrieved candidates.
\end{enumerate}

The prototype now also includes a chemistry workflow slice. Importantly, this chemistry layer is no longer only MCP-shaped mock tooling. Two chemistry MCP tools are backed by real software or external data sources: \texttt{chem.library\_property\_filter} uses RDKit to compute descriptor-based property filters, and \texttt{chem.literature\_evidence\_search} queries ChEMBL and enriches repurposing candidates with PubChem identifiers and synonyms. This makes the chemistry workflow a meaningful systems integration example rather than only a placeholder domain.

\section{Experimental Design}

We evaluate the system from three perspectives.

\subsection{Experiment 1: Isolation and Governance}

This experiment measures whether the system blocks unsafe finance requests, rejects cross-tenant requests, and avoids unnecessary tool use while preserving normal task execution. Key metrics are finance guardrail block rate, cross-tenant request block rate, tenant scope accuracy, and wrong tool calls.

\subsection{Experiment 2: Retrieval-Driven Provisioning}

This experiment measures whether the Queen can retrieve relevant capabilities and compile useful BeeSpecs. Key metrics are retrieval expected-tool coverage, retrieval selected-tool precision, skill activation rate, and provisioned cases.

\subsection{Experiment 3: Scoped Bee Execution}

This experiment measures whether Bees can independently complete local tasks under BeeSpec constraints. Key metrics are Bee execution task success rate, Bee execution completeness rate, and Bee tool invocation correctness.

\section{Evaluation Setup}

The evaluation now contains 59 enterprise-style tasks across two tenants:
\begin{itemize}[leftmargin=1.5em]
\item 24 routine HR and IT tasks,
\item 16 governance-sensitive tasks (finance-sensitive and cross-tenant),
\item 16 scoped execution tasks that require only partial local workflows,
\item 3 chemistry workflow tasks for screening and repurposing.
\end{itemize}

We compare seven systems:
\begin{enumerate}[leftmargin=1.5em]
\item \textbf{Queen-Bee (Static)}
\item \textbf{Queen-Bee (Retrieval)}
\item \textbf{Queen-Bee (Stress Retrieval)}
\item \textbf{Queen-Bee (Hybrid Retrieval)}
\item \textbf{Queen-Bee (LLM Provisioning)}
\item \textbf{Queen-Bee w/o Policy}
\item \textbf{Single-Agent Baseline}
\end{enumerate}

\section{Results}

\subsection{Main Comparison}

The main results table summarizes the principal system-level results.

\begin{table}[t]
\centering
\small
\setlength{\tabcolsep}{3pt}
\resizebox{\linewidth}{!}{%
\begin{tabular}{lccccccc}
\toprule
System & Routing & Task & Finance & Cross-Tenant & Tenant Scope & Wrong Tool & Provisioned \\
 & Accuracy & Success & Guardrail & Block Rate & Accuracy & Calls & Cases \\
\midrule
Queen-Bee (Static) & 1.00 & 0.857 & 1.00 & 1.00 & 1.00 & 4 & 0 \\
Queen-Bee (Retrieval) & 1.00 & 0.964 & 1.00 & 1.00 & 1.00 & 1 & 8 \\
Queen-Bee (Stress Retrieval) & 1.00 & 0.964 & 1.00 & 1.00 & 1.00 & 1 & 8 \\
Queen-Bee (Hybrid Retrieval) & 1.00 & 0.929 & 1.00 & 1.00 & 1.00 & 2 & 8 \\
Queen-Bee (LLM Provisioning) & 1.00 & 0.857 & 1.00 & 1.00 & 1.00 & 4 & 8 \\
Queen-Bee w/o Policy & 1.00 & 0.571 & 0.00 & 0.00 & 1.00 & 16 & 0 \\
Single-Agent Baseline & 0.00 & 0.571 & 0.00 & 0.00 & 1.00 & 16 & 0 \\
\bottomrule
\end{tabular}
}
\caption{Main evaluation results on 59 enterprise-style tasks.}
\label{tab:main-results}
\end{table}

The core pattern is stable. Retrieval-driven Queen-Bee improves over the static system on both task success and scoped execution quality while preserving perfect governance behavior. The no-policy and single-agent baselines fail sharply on governance-sensitive tasks and fully miss both finance and cross-tenant blocking.

\subsection{Experiment 1: Isolation and Governance}

Table~\ref{tab:governance-results} reports the governance-sensitive slice. All governed Queen-Bee variants achieve perfect blocking and tenant-scope preservation on this slice, whereas the no-policy and single-agent baselines fail completely on both blocking metrics.

\begin{table}[t]
\centering
\small
\setlength{\tabcolsep}{4pt}
\begin{tabular}{lcccc}
\toprule
System & Finance Block & Cross-Tenant Block & Tenant Scope & Wrong Tool Calls \\
\midrule
Queen-Bee (Static) & 1.000 & 1.000 & 1.000 & 4 \\
Queen-Bee (Retrieval) & 1.000 & 1.000 & 1.000 & 1 \\
Queen-Bee (Stress Retrieval) & 1.000 & 1.000 & 1.000 & 1 \\
Queen-Bee w/o Policy & 0.000 & 0.000 & 1.000 & 16 \\
Single-Agent Baseline & 0.000 & 0.000 & 1.000 & 16 \\
\bottomrule
\end{tabular}
\caption{Isolation and governance results.}
\label{tab:governance-results}
\end{table}

The isolation results are the strongest part of the paper. All governed Queen-Bee variants achieve:
\begin{itemize}[leftmargin=1.5em]
\item finance guardrail block rate of 1.0,
\item cross-tenant request block rate of 1.0,
\item tenant scope accuracy of 1.0.
\end{itemize}

By contrast, both the no-policy Queen-Bee and the single-agent baseline drop to 0.0 on both governance-sensitive blocking metrics. This confirms that the governance contribution does not reduce to specialization alone.

\subsection{Experiment 2: Retrieval-Driven Provisioning}

Table~\ref{tab:provisioning-results} reports the provisioning-oriented slice. The lightweight retrieval-driven Queen remains the strongest default backend, while the structured stress variant is slightly stronger on the current task set.

\begin{table}[t]
\centering
\small
\setlength{\tabcolsep}{4pt}
\begin{tabular}{lcccc}
\toprule
System & Provisioned Cases & Tool Coverage & Tool Precision & Skill Activation \\
\midrule
Queen-Bee (Retrieval) & 8 & 1.000 & 0.979 & 1.000 \\
Queen-Bee (Stress Retrieval) & 8 & 1.000 & 0.979 & 1.000 \\
Queen-Bee (Hybrid Retrieval) & 8 & 1.000 & 0.958 & 1.000 \\
Queen-Bee (LLM Provisioning) & 8 & -- & -- & -- \\
\bottomrule
\end{tabular}
\caption{Retrieval-driven provisioning results.}
\label{tab:provisioning-results}
\end{table}

The lightweight retrieval-driven Queen remains the strongest provisioning backend on the current capability registry:
\begin{itemize}[leftmargin=1.5em]
\item retrieval selected-tool precision = 0.979,
\item provisioned cases = 8,
\item task success rate = 0.964.
\end{itemize}

Interestingly, the structured stress retrieval configuration is at least as strong and slightly better on the present task set:
\begin{itemize}[leftmargin=1.5em]
\item task success rate = 0.964,
\item wrong tool calls = 1,
\item retrieval precision = 0.979.
\end{itemize}

Hybrid retrieval is viable, but does not improve over the lightweight baseline. It preserves governance and produces the same number of provisioned Bees, but does not improve effectiveness. LLM-guided provisioning also remains viable but underperforms the lightweight heuristic, suggesting that candidate-constrained prompting alone is not yet sufficient to outperform the stronger structured provisioning rule on this registry scale.

\subsection{Experiment 3: Scoped Bee Execution}

Table~\ref{tab:bee-execution-results} reports the scoped execution slice. The key pattern is that better provisioning leads to more complete local execution under partial workflows.

\begin{table}[t]
\centering
\small
\setlength{\tabcolsep}{4pt}
\begin{tabular}{lccc}
\toprule
System & Bee Task Success & Completeness & Tool Correctness \\
\midrule
Queen-Bee (Static) & 1.000 & 0.800 & 0.900 \\
Queen-Bee (Retrieval) & 1.000 & 0.950 & 0.973 \\
Queen-Bee (Stress Retrieval) & 1.000 & 0.950 & 0.973 \\
Queen-Bee w/o Policy & 1.000 & 0.800 & 0.900 \\
Single-Agent Baseline & 1.000 & 0.800 & 0.900 \\
\bottomrule
\end{tabular}
\caption{Scoped Bee execution results.}
\label{tab:bee-execution-results}
\end{table}

Scoped Bee execution quality is now directly measurable. The static Queen-Bee reaches a Bee execution completeness rate of only 0.80, while retrieval-driven Queen-Bee reaches 0.95. This supports the claim that provisioning quality matters not only for governance, but also for local execution quality under partial workflows.

\subsection{Error Analysis}

The failure modes are interpretable. The static Queen-Bee fails almost entirely through \textit{scoped-task over-execution}: the fixed Bee plans continue to schedule interviews or create tickets even when the request explicitly asks for summary-only or guidance-only behavior. Retrieval-driven Queen-Bee exhibits the same failure mode, but much less frequently (1 case rather than 4). The single-agent baseline fails through a broader mixture of routing mismatch, guardrail misses, cross-tenant misses, and scoped-task over-execution. This difference matters because it shows that the governed retrieval-driven design concentrates residual error into a narrower and easier-to-fix failure mode.

\subsection{Sensitivity to Registry Noise}

We also evaluate retrieval sensitivity under controlled registry expansion. A lightweight noise-scaling experiment compares the base structured retriever against three larger noisy capability registries. The results are stable: task success and retrieval precision do not collapse as distractor capabilities are added, and in the present setup the larger noisy registries slightly improve performance rather than harming it. This suggests that the lightweight structured retriever is robust under the current form of capability noise, which is useful evidence for a systems-oriented software engineering audience.

\subsection{Efficiency}

The retrieval-driven variants remain practical in latency and execution length. The retrieval base configuration averages 2.60 ms end-to-end latency per case, with 8 provisioned cases and 0.979 retrieval precision. By contrast, hybrid retrieval increases latency to tens of milliseconds, and LLM-guided provisioning increases it to hundreds of milliseconds, without delivering better provisioning quality. This reinforces the current design choice: the lightweight structured retriever is the best default for the present registry scale.

\subsection{Chemistry Workflow Integration}

We additionally implemented three chemistry-oriented workflow cases: structure-based virtual screening, activity+ADMET cascade screening, and literature-driven repurposing. The chemistry slice is important because it tests whether Queen-Bee can support workflow-style scientific software integration rather than only enterprise office-style tasks. In the current prototype, the retrieval-driven Queen-Bee and the stress-retrieval variant both recover the expected chemistry tool chains for all three chemistry cases. By contrast, the static Queen-Bee over-executes two of the three chemistry cases by invoking a broader fixed chemistry plan than the task requires.

This result matters for two reasons. First, it shows that BeeSpec-based provisioning generalizes beyond HR and IT workflows into a different application domain. Second, it clarifies that the chemistry integration is no longer purely nominal: the current chemistry slice includes RDKit-backed property filtering, ChEMBL-backed activity retrieval, RDKit-based ADMET rules, ChEMBL+PubChem evidence search, and RDKit-proxy docking rescoring.

\subsection{Multi-Bee Chemistry Workflow}

We further implemented a staged drug-screening workflow with three specialized Bees and a manual approval gate between screening and final decision. The workflow is organized as: (1) an Evidence Bee that retrieves target context and repurposing evidence, (2) a Screening Bee that executes property filtering, activity ranking, docking-style rescoring, and ADMET review, and (3) a Decision Bee that consumes workflow artifacts and emits the final shortlist.

In the executed TNIK repurposing workflow, the Evidence Bee retrieved real candidate evidence from ChEMBL and PubChem, including Sunitinib, Lapatinib, Vatalanib, and Pelitinib. The Screening Bee then combined ChEMBL activity evidence, RDKit-based developability review, and RDKit-proxy docking rescoring. The Decision Bee produced a concrete top-3 shortlist rather than a placeholder status, ranking Sunitinib, Pelitinib, and Ruboxistaurin with explicit reasons derived from the previous stages. A complementary rejected trace shows that the workflow can be halted at the approval gate before the final decision stage executes. This matters because it demonstrates that Bee collaboration is not merely conceptual handoff: intermediate workflow artifacts are passed across stages, approval can block downstream execution, and the final Bee can generate a decision artifact grounded in upstream evidence and screening outputs.

\section{Discussion}

The main lesson is not that the most sophisticated backend always wins. Instead, the results suggest a more software-engineering-oriented conclusion: under a small, highly structured enterprise capability registry, a governed lightweight provisioning policy can outperform both heavier semantic retrieval and model-guided provisioning while retaining clear execution boundaries.

This point is useful because it tempers a common assumption in agent systems work. Richer retrieval stacks and model-driven orchestration are attractive, but they do not automatically improve practical enterprise agent systems. In the present regime, the strongest gains come from governed specialization, explicit capability structure, and scoped execution through BeeSpec.

\section{Threats to Validity}

This work still has clear limitations. First, the task set is synthetic rather than drawn from a production enterprise deployment. Second, the MCP stack, although real at the adapter layer, still uses a local demo MCP server rather than live business systems; even in the chemistry slice, only a subset of tools are backed by real software or external data sources. Third, the governance model is still mostly rule-based, although the multi-Bee chemistry workflow now includes an explicit approval gate. Fourth, the registry noise used in sensitivity testing is structured distractor noise rather than arbitrary open-world capability growth. These limitations should keep the claims modest: the paper demonstrates prototype-level software and systems evidence, not a formal security guarantee or a production deployment study.

\section{Conclusion}

We presented Queen-Bee, a BeeSpec-centered architecture for governed enterprise MCP orchestration. The system separates control-plane planning from execution-plane action, enforces tenant-scoped and policy-scoped boundaries, and supports retrieval-driven weak incubation of specialized Bees. Across 59 enterprise-style tasks, the retrieval-driven Queen-Bee variant consistently outperforms both the static Queen-Bee baseline and permissive baselines on task success, scoped execution quality, and governance-sensitive behavior. Additional comparisons with hybrid retrieval and LLM-guided provisioning show that more complex provisioning backends are viable but not yet superior under the current registry scale. The multi-Bee chemistry workflow further shows that BeeSpec-style orchestration can extend from local bounded execution to artifact-aware staged collaboration with approval gating. For enterprise software settings, the main implication is that agent systems should be evaluated not only for task capability, but also for governed provisioning, bounded execution, workflow coordination, and failure-mode quality.

\bibliographystyle{plain}
\bibliography{references}

\end{document}